# The RKKY coupling in diluted magnetic semiconductors

A. M. Werpachowska[(1,2)] and Z. Wilamowski[(1)]

[(1)]*Institute of Physics PAS, 02-668 Warsaw, Poland*

[(2)]*College of Science by the Polish Academy of Sciences, Warsaw, Poland*

This paper is an attempt to modify the classic Ruderman-Kittel-Kasuya-Yosida (RKKY) model to allow the analysis of the magnetic resonance measurements. In our calculations, we follow the treatment of the original authors of the RKKY model but include the finite band splitting, $\Delta$, as a phenomenological parameter. The RKKY exchange is not anymore of Heisenberg type and an anisotropy induced by the direction of carrier magnetization occurs.

## I. INTRODUCTION

The RKKY coupling [1,2], i.e. the exchange interaction between localized core spins mediated by metallic electron gas, has been known for 50 years as the basic interaction in metallic ferromagnets. The oscillatory character of the RKKY coupling causes a spin glass behaviour in diluted magnetic metals. It rules the interlayer coupling in magnetic layered structures. As it has been shown for the last few years, the RKKY interaction is also the dominant spin interaction in diluted ferromagnetic semiconductors [3,4]. However, in the case of semiconductors, strictly speaking of the semi-metallic phase of semiconductors, the Fermi energy, $E_F$, is small as compared to classical metals and comparable to the exchange spin splitting of the conduction band, $\Delta$. The small Fermi energy causes a saturation of spin polarization of carrier spins and some other new effects [5].

In this paper we discuss consequences of the spin splitting on the RKKY coupling. The investigation of analytical expressions for the distance dependence of the exchange coupling shows that in the presence of spin splitting the RKKY exchange is not anymore of Heisenberg type but leads to magnetization-induced anisotropy. With an increase of $\Delta$, which we treat as a phenomenological parameter, the RKKY coupling evolves from the classical Heisenberg coupling, via anisotropic interaction to the Ising coupling. Moreover, the spin splitting leads to the occurrence of various contributions to the RKKY coupling which are characterized by different distance dependencies and various characteristic lengths.

## II. MAGNETIC RESONANCE IN (GA,MN)AS

The formation of magnetic order in semi-metallic (Ga,Mn)As is well described by the Mean Field Approach (MFA) models [6]. It is now commonly accepted that the p-d exchange is responsible for the hole mediated exchange. Within the MFA, the Zener [7], RKKY [1,2] and Dietl [3,4] models are equivalent. All of them permit a good estimation of the critical temperature. The Dietl model, which attributes the p-d energy to the carrier spins and takes into consideration the details of the valence band structure, additionally allows for the estimation of the magnetic anisotropy. Unfortunately, the aforementioned models calculate the energy of the ground state of the system but they do not analyse the elementary excitations. Therefore, they cannot be directly applied to the discussion of the magnetic resonance.

Studies of the magnetic resonance show that are two types of the magnetic resonance observed in (Ga,Mn)As [8,9]. None of them, however, can be attributed to any magnetic resonance already observed in similar systems [10-12]. Up till now, none of them is satisfactorily explained. In particular, ferromagnetic resonance in the semi-insulating (Ga,Mn)As is characterized by the resonance frequency which matches precisely the manganese spins' resonance frequency. The absence of any influence of the hole spins remains unexplained. The resonance in the semi-metallic (Ga,Mn)As is of a very different character. It is characterized by a large anisotropy, a complex structure which can be attributed to the spin wave resonance and $g$-factor considerably different from that corresponding to the $g$-factor of the Mn spins. These properties allow to conclude that the resonance corresponds to the ferromagnetic mode of the ferrimagnetic resonance [8,9]. It suggests that the carrier spins are ordered and form a macroscopic magnetic moment. Till now that problem has not been considered by any theoretical model. The exchange correction within the Fermi liquid approach is the only effect discussed by Dietl et al. [3,4]. All of the approaches discussed postulate paramagnetic properties of the carrier spins.

A quantitative description of the ferrimagnetic resonance requires the solution of the equation of motion of two interacting spin subsystems. The precise definition of the tensor components describing spins' coupling is of a crucial importance. In particular, the two cases: the Heisenberg exchange and the effective field lead to very different types of precession. For example, according to the RKKY model, which treats the carriers as a paramagnetic medium and postulates a Heisenberg coupling between the localized spins, the resonance in the magnetically ordered semi-metallic (Ga,Mn)As should correspond to the isotropic resonance with $g = 2$. On the other hand, according to the Zener model, which predicts a huge mean p-d exchange field acting on each spin subsystem, the two resonances corresponding to Mn and hole spins are expected at very high frequencies. None of the models corresponds to the experimental observations.

## III. THE EFFECT OF THE BAND SPLITTING ON THE RKKY INTERACTION

The aim of this paper is a critical study of the RKKY interaction and an attempt to modify the classic RKKY model to allow for the analysis of the experimental data of the magnetic resonance measurements. In our calculations, we follow the treatment of the original authors of the RKKY model, the difference being that we assume a finite band splitting, $\Delta$. We do not specify whether this splitting comes from the spontaneous magnetization of the local spins' subsystem, is the giant spin splitting typical for the DMS's, $\Delta = \{N_0\beta\}x_{Mn}\langle S\rangle$, or finally, whether it is caused by the external magnetic field, $\Delta_B = g\mu_B B$. We obtain the analytical formula for the direction dependent RKKY range functions, which are presented in the Appendix.

Fig. 1 shows an example of the dependence of the exchange integrals on the distance for $\Delta = 0.25\ E_{F0}$, here $E_{F0}$ is the Fermi energy in the absence of the spin splitting. All of them are of oscillating character but their amplitudes and oscillation frequencies are different. As a consequence of one direction being privileged by the splitting $\Delta$, a magnetization-induced anisotropy arises. As opposed to the classic RKKY coupling, which has the form of a strictly Heisenberg-like interaction, the RKKY tensor ($J_{RRKY}$) components are not all equal in our calculations. The $J_{zz}$ component corresponds to the direction parallel to the conduction band magnetization and the components $J_{xx} = J_{yy}$ correspond to the perpendicular direction.

The $J_{zz}(r)$ component is a sum of two components (see Eq. 1 in the Appendix). Each of them is the classic RKKY function for different Fermi vectors $k_{F\uparrow}$ and $k_{F\downarrow}$ (corresponding to the spin up and spin down subbands). Due to differing frequencies, the exchange interaction range $\lambda = \pi/4k_F$ is different for each of the two contributions. The inset to Fig. 1a shows the dependence of the distance $r_1$, for which the $J_{zz}$ component reaches zero value for the first time. The distance for the majority spin subband systematically decreases, while the distance to the first zero increases and diverges for a complete polarization.

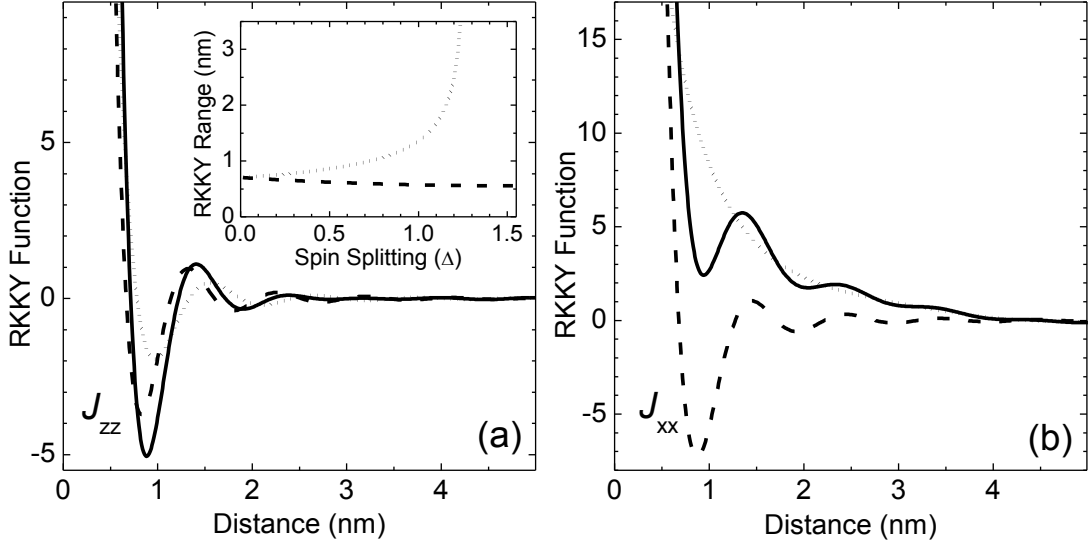

Fig.1 The components of the RKKY function as a function of the distance calculated for a simple parabolic band with the electron concentration $n_c = 10^{27}$ m$^{-3}$: a) The solid line represents the $J_{zz}(r)$ component, the dashed and the dotted line stand for the contributions for the spin up and spin down subbands. The insert shows the dependence of the characteristic RKKY range on the spin splitting, $\Delta$. The positions of the first zero of $J(r)$ functions, $r_1$ are plotted. b) The transverse component $J_{xx}(r)$ is plotted by the solid line. The dashed and the dotted line correspond to the two contributions. The long range contribution oscillates in the range of large distances beyond the figure frame.

It is notable that the $J_{zz}$ component (Fig. 1a) does not vanish for the half-metal case, when only one of the spin subbands is occupied (for $\Delta$ larger than the Fermi energy). In this regime, the exchange interaction is carried by the total polarized charge density of the electron gas (Friedel oscillations). Generally, we can treat the classic RKKY interaction as a sum of the two Friedel contributions. In the $\Delta \to 0$ limit the contributions are precisely equal and the total charge density oscillations vanish, while the spin polarization (RKKY) oscillations remain. For the finite spin splitting the contributions are not equal anymore. Finally for $\Delta > 0$, when only the majority spin subband is occupied, the exchange is mediated by the Friedel oscillations only.

The $J_{xx}(r)$ component (see Eq. 2 in the Appendix) may be expressed as the sum of two qualitatively different contributions (Fig. 1b). One of them, shown in Fig. 1b by the dashed line, is the modified RKKY function with the oscillations corresponding to the sum of the Fermi vectors. The amplitude of this contribution gets smaller as $\Delta$ increases and vanishes in the half-metal regime. The second contribution shown in Fig. 1b by the dotted line oscillates with the frequency corresponding to the difference between the Fermi vectors. It is therefore characterized by the long characteristic range. The amplitude of this contribution grows with $\Delta$ and in the half-metal regime decreases with the further increase of $\Delta$.

The amplitudes of particular contributions are better seen in Fig. 2, where the mean exchange field, $\Xi$, proportional to the volume integral of the $J(r)$ function is plotted as a function of $\Delta$. For the whole range of $\Delta$, the longitudinal component of the exchange field, $\Xi_{zz}$, is bigger than the transverse one, $\Xi_{xx}$. It means that the magnetization of the local spin has the tendency to be oriented parallel to the $z$ direction, as defined by the magnetization of the carrier spins. This tendency vanishes with vanishing $\Delta$. Apart from the listed dependencies of the contributions on the spin splitting, we see the onset of the magnetization-induced anisotropy. The anisotropy, $D = \Xi_{zz} - \Xi_{xx}$, corresponds to the magnetization-induced anisotropy. The coupling between the local spins evolves with $\Delta$, from the pure Heisenberg coupling for $\Delta = 0$, via anisotropic to the Ising coupling in the large $\Delta$ limit, where the transverse component vanishes. In this regime, the exchange field, as seen by a random local spin, is parallel to the electron gas magnetization. In this sense, the Ising form of the exchange coupling between the localized spins corresponds to the Zener model, where the p-d coupling between both spin subsystems also has only the component parallel to the $z$ axis. The both models, Zener and RKKY neglect the transverse components. However, the Zener approach explains the lack of influence of the perpendicular components by the postulated averaging of the perpendicular spin components (random phases of spin precession), while the lack of the transverse component of RKKY exchange is the consequence of the half-metal character and the large spin splitting of the carrier band.

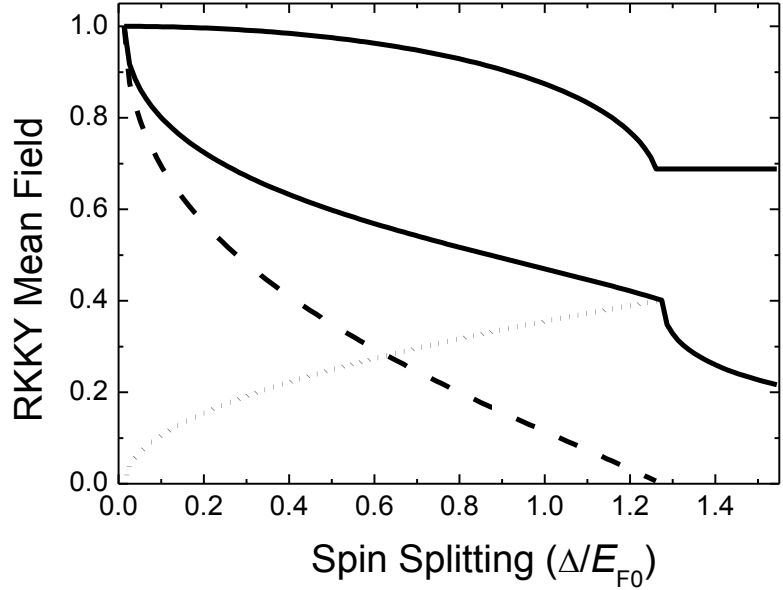

Fig.2 The dependence of the two RKKY mean field components on the spin splitting $\Delta$. The two contributions to the transverse component $\Xi_{xx}$ are marked by the dotted and dashed lines.

## IV. CONCLUSIONS

Contrary to the intuition given by the Zener model, which suggests an (anti)parallel orientation of both magnetic moments, tilting of the directions of magnetizations is not that energetically expensive. Only in the limit of large $\Delta$, within the range of the Ising exchange, the energy of the magnetization deviation as predicted by the RKKY coupling and that estimated by the Zener model, are similar. For small value of $\Delta$, however, the energy gain for the collinear magnetizations is not as significant as suggested by the Zener model.

The anisotropic part of the RKKY interaction, which results from the band spin splitting, is not very important when discussing the ground state of the system, i.e. when the directions of the Mn and the carrier spins are parallel. Its importance grows, e.g. when the carrier magnetization is tilted by an anisotropy field, or under an external field when the precession angles of both magnetizations are different. It may be a possible explanation for the quantitative description of the hysteresis loops. Magnetic anisotropy parameters evaluated from the magnetic resonance studies are not able to describe the hysteresis loops observed in the transport, Kerr and SQUID measurements.

The described character of the exchange integrals should also lead to a peculiar spin wave dispersion and domain wall structure. The appearance of several contributions to the range functions components with the different characteristic ranges results in the non-parabolic spin wave dispersion. While the occurrence of the long-range component of the RKKY coupling may lead to the big magnetic stiffness and consequently to the high energy of the domain wall. However, this kind of anisotropy has no direct influence on the observed anisotropy of the spin wave dispersion.

## V. ACKNOWLEDGEMENT

This work was supported by PBZ-KBN-044/P03/2001.

## APPENDIX

The function $J_{zz}$ is given by an analytical formula

$$J_{zz}(r) = \{N_0\alpha\}^2 \left( \frac{m^*(\sin(2k_{F\uparrow}r) - 2k_{F\uparrow}r\cos(2k_{F\uparrow}r))}{4\hbar^2\pi^3 r^4} + \frac{m^*(\sin(2k_{F\downarrow}r) - 2k_{F\downarrow}r\cos(2k_{F\downarrow}r))}{4\hbar^2\pi^3 r^4} \right), \quad (1)$$

where $m^*$ is the effective mass, $\hbar$ is the Planck constant over $2\pi$ and $\{N_0\alpha\}$ is the exchange constant.

The formula for $J_{xx} = J_{yy}$ function is

$$\begin{aligned}
J_{xx}(r) = &\frac{m^*\{N_0\alpha\}^2}{8\hbar^2\pi^3 r^4} \Big( (k_{F\uparrow} + k_{F\downarrow})r\big((k_{F\uparrow} - k_{F\downarrow})^2 r^2 - 2\big)\cos\big((k_{F\uparrow} + k_{F\downarrow})r\big) \\
&+ \sqrt{k_{F\uparrow}^2 - k_{F\downarrow}^2}\, r\big(2 - (k_{F\uparrow}^2 - k_{F\downarrow}^2)r^2\big)\cos\big(\sqrt{k_{F\uparrow}^2 - k_{F\downarrow}^2}\, r\big) \\
&+ \big(2 + (k_{F\uparrow} - k_{F\downarrow})^2 r^2\big)\sin\big((k_{F\uparrow} + k_{F\downarrow})r\big) - \big(2 + (k_{F\uparrow}^2 - k_{F\downarrow}^2)r^2\big)\sin\big(\sqrt{k_{F\uparrow}^2 - k_{F\downarrow}^2}\, r\big) \\
&+ (k_{F\uparrow}^2 - k_{F\downarrow}^2)r^4\Big(\mathrm{Si}\big((k_{F\uparrow} + k_{F\downarrow})r\big) - \mathrm{Si}\big(\sqrt{k_{F\uparrow}^2 - k_{F\downarrow}^2}\, r\big)\Big)\Big) \\
&+ \frac{m^*\{N_0\alpha\}^2}{4\hbar^2\pi^2 r^3}\Big((k_{F\uparrow}^2 - k_{F\downarrow}^2)r H_0\big(\sqrt{2}\sqrt{k_{F\uparrow}^2 - k_{F\downarrow}^2}\, r\big) - \sqrt{2}\sqrt{k_{F\uparrow}^2 - k_{F\downarrow}^2}\, H_1\big(\sqrt{2}\sqrt{k_{F\uparrow}^2 - k_{F\downarrow}^2}\, r\big)\Big)
\end{aligned} \quad (2)$$

where

$$\mathrm{Si}(z) = \int_0^z \frac{\sin t}{t}\, dt$$

and the Struve function $H_n(z)$ [13, p. 496] for integer $n$ satisfies the differential equation

$$z^2 y'' + z y' + (z^2 - n^2) y = \frac{2}{\pi} \frac{z^{n+1}}{(2n-1)!!}.$$